\documentclass[11pt,a4paper]{article}
\usepackage{amsmath,amssymb}
\usepackage{epsfig,graphicx}
\usepackage{bm}

\begin{document}

\title{\bf Troubles of  describing multiple pion production in chiral
dynamics.}
\author{N.~N.~Achasov$^a$\footnote{{\bf e-mail}: achasov@math.nsc.ru}, \underline{A.~A.~Kozhevnikov}
$^{a,b}$\footnote{{\bf e-mail}: kozhev@math.nsc.ru} \\
$^a$
\small{\em Laboratory of Theoretical Physics, S.~L.~Sobolev Institute for Mathematics} \\
\small{\em 630090, Novosibirsk, Russian Federation} \\
$^b$
\small{\em  Novosibirsk State University} \\
\small{\em 630090, Novosibirsk, Russian Federation} }
\date{}
\maketitle

\begin{abstract}
Generalized Hidden Local Symmetry (GHLS) model as the chiral model
of pseudoscalar, vector, and axial vector mesons and their
interactions containing also the couplings of strongly interacting
particles with electroweak gauge bosons,  is confronted with the
ALEPH data on the decay $\tau^-\to\pi^-\pi^-\pi^+\nu_\tau$ and
BABAR and CMD data on the reaction
$e^+e^-\to\pi^+\pi^-\pi^+\pi^-$. It is shown that both the
invariant mass spectrum of final pions in $\tau$ decay  calculated
in the GHLS framework with the single $a_1(1260)$ resonance and
the cross section $e^+e^-\to\pi^+\pi^-\pi^+\pi^-$ calculated in
the above framework with the single $\rho(770)$ resonance,
disagree with the experimental data. The modifications of GHLS
model based on inclusion of two additional heavier axial vector
mesons $a_1^\prime$, $a_1^{\prime\prime}$ in the $\tau$ decay and
the vector mesons $\rho^\prime$, $\rho^{\prime\prime}$ in
$e^+e^-\to\pi^+\pi^-\pi^+\pi^-$ are shown to be necessary for the
good description of the above data.
\end{abstract}

{\bf 1. Introduction.} The theory aimed at describing low energy
hadron processes should be formulated in terms of effective
colorless degrees of freedom introduced on the basis of
spontaneously broken approximate chiral symmetry $SU(3)_L\times
SU(3)_R$. This is the symmetry of QCD Lagrangian
\begin{eqnarray}
{\cal L}_{\rm QCD}&=&-\frac{1}{4}\left(\partial_\mu
G^a_\nu-\partial_\nu G^a_\mu+gf_{abc}G^b_\mu G^c_\nu\right)^2
+\sum_q\bar q\left[\gamma_\mu\left(i\partial_\mu-
g\frac{\lambda^a}{2}G_{a\mu}\right)- m_q\right]q,
\end{eqnarray}
relative independent rotations of right and left fields of
approximately massless $u,d,s$ quarks:
\begin{eqnarray}
q_L\equiv\frac{1+\gamma_5}{2}q&\rightarrow& V_Lq_L\mbox{,
}q_R\equiv\frac{1-\gamma_5}{2}q\rightarrow V_Rq_R,
\end{eqnarray}
where $V_{L,R}\in SU(3)_{L,R}$. The pattern of the spontaneous
breaking is $SU(3)_L\times SU(3)_R\Rightarrow SU(3)_{L+R}$.
According to the Goldstone theorem, spontaneous breaking of global
symmetry results in appearance of massless fields. In our case
they are light $J^P=0^-$ mesons $\pi^+$, $\pi^-$, $\pi^0$, $K^+$,
$K^0$, $K^-$, $\bar K^0$, $\eta$. The transformation law
$U\rightarrow V_LUV^\dagger_R$ where
$U=\exp\left(i{\bm\Phi}\sqrt{2}/f_\pi\right)$, and
\begin{eqnarray}{\bm\Phi}&=&\left(
\begin{array}{ccc}
\frac{\pi^0}{\sqrt{2}}+\frac{\eta}{\sqrt{6}}&\pi^+&K^+\\
\pi^-&-\frac{\pi^0}{\sqrt{2}}+\frac{\eta}{\sqrt{6}}&K^0\\ K^-&\bar
K^0&-\frac{2\eta}{\sqrt{6}}
\end{array}\right),\;
\end{eqnarray}
fixes the Lagrangian of interacting Goldstone mesons:
$$
{\cal L}_{\rm GB}=\frac{f^2_\pi}{4}\mbox{Sp}\left(\partial_\mu
U\partial_\mu U^\dagger\right)+\cdots.$$ Dots mean the terms with
higher derivatives. Upon adding the term $\propto
m^2_\pi\mbox{Sp}(U+U^\dagger)$ which explicitly breaks chiral
symmetry, Goldstone bosons become massive.

Pseudoscalar mesons are produced  via intermediate  vector and
axial resonances, hence one should  include vector and axial
vector mesons in a chiral invariant way. There are a number of
chiral models of pseudoscalar, vector, and axial mesons and their
interaction \cite{meissner88}.  The problem of testing chiral
models of the vector meson interactions with  Goldstone bosons is
acute. The present report is devoted to reviewing the attempts to
confront one of the chiral models,  the  Generalized Hidden Local
Symmetry (GHLS) model \cite{bando85,bando88a,bando88}, with the
data on the decay $\tau^-\to\pi^+\pi^-\pi^-\nu_\tau$
\cite{aleph05} in the axial vector channel, and the data on the
reaction $e^+e^-\to\pi^+\pi^-\pi^+\pi^-$ \cite{cmd2,babar} in the
vector channel, both in the state with the isospin one.

{\bf 2. Generalized Hidden Local Symmetry Model Lagrangian.} The
Generalized Hidden Local Symmetry (GHLS) model
\cite{bando85,bando88a,bando88} as the chiral model based on
nonlinear realization of chiral symmetry,  is of a special
interest because some interesting two- and three-particle decays,
for example, $\rho^0\to\pi^+\pi^-$ and $\omega\to\pi^+\pi^-\pi^0$,
were analyzed in  its framework  \cite{bando88}. "Hidden" means
that if $U=\xi^\dag_L\xi_R$ then the transformation law
$\xi_{L,R}\to h\xi_{L,R}g^\dag_{L,R}$ implies one  $U\to
g_LUg^\dag_R$ where $h$ transforms vector meson fields in a
gauge-like manner as $V_\mu\to hV_\mu h^\dag-i\partial_\mu
hh^\dag$. "Generalized hidden" means that axial vector mesons are
included. One of the virtues of GHLS model is that the sector of
electroweak interactions is introduced in such a way that the low
energy relations in the sector of strong interactions are not
violated upon inclusion of photons and electroweak gauge bosons.
The GHLS lagrangian  includes pseudoscalar, vector, and axial
vector fields $\xi$, $V_\mu$, and $A_\mu$, respectively. In the
gauge $\xi_M=1$, $\xi^\dagger_L=\xi_R=\xi$ and after rotating away
the axial vector-$\pi$ mixing by choosing
\begin{equation}
A_\mu=a_\mu-\frac{b_0c_0}{g(b_0+c_0)}A_{(\xi)\mu},\label{physa1}\end{equation}
where $a_\mu$ is $a_1$ meson field, $g$ is the coupling constant
to be related to $g_{\rho\pi\pi}$, and
\begin{equation}
A_{(\xi)\mu}=\frac{\partial_\mu\xi^\dag\xi-\partial_\mu\xi\xi^\dag}{2i},\label{Axi}\end{equation}
the relevant terms corresponding to  strong interactions look like
\begin{eqnarray}
{\cal L}_{\rm strong}&=&a_0f^{(0)2}_\pi{\rm
Tr}\left(\frac{\partial_\mu\xi^\dagger\xi+\partial_\mu\xi\xi^\dagger}{2i}-gV_\mu\right)^2+
f^{(0)2}_\pi\left(d_0+\frac{b_0c_0}{b_0+c_0}\right){\rm
Tr}A^2_{(\xi)\mu}+\nonumber\\&&(b_0+c_0)f^{(0)2}_\pi g^2{\rm
Tr}a^2_\mu+d_0f^{(0)2}_\pi{\rm Tr}A^2_{(\xi)\mu} -\frac{1}{2}{\rm
Tr}\left(F^{(V)2}_{\mu\nu}+F^{(A)2}_{\mu\nu}\right)-\nonumber\\&&i\alpha_4g{\rm
Tr}[A_\mu,A_\nu]F^{(V)}_{\mu\nu}+2i\alpha_5{\rm
Tr}\left(\left[A_{(\xi)\mu},A_\nu\right]
+g[A_\mu,A_\nu]\right)F^{(V)}_{\mu\nu}.
\label{lghls}\end{eqnarray}
The lagrangian contains a number of free parameters
$a_0,b_0,c_0,d_0,\alpha_4,\alpha_5$. The terms with free
parameters  $\alpha_{4,5}$ are necessary for cancelation of
momentum dependence in the $\rho\pi\pi$ vertex. They are chosen in
accord with Refs.~\cite{bando88a,bando88} in such a way that among
the terms with higher derivatives those with
$\alpha_1,\alpha_2,\alpha_3$ are set to zero, and only the
$\alpha_{4,5,6}$ terms are included, with the additional
assumption $\alpha_5=\alpha_6$ about the arbitrary constants
multiplying the lagrangian terms. The remaining ones $\alpha_4$
and $\alpha_5$ should be related like
\begin{equation}
\alpha_4=1-\frac{2\alpha_5c_0}{b_0},\label{alpha45}\end{equation}
in order to provide the desired  cancelation. The notations,
assuming the restriction to the sector of the non-strange mesons,
are
\begin{eqnarray}
F^{(V)}_{\mu\nu}&=&\partial_\mu V_\nu-\partial_\nu
V_\mu-ig[V_\mu,V_\nu]-ig[A_\mu,A_\nu],\nonumber\\
F^{(A)}_{\mu\nu}&=&\partial_\mu A_\nu-\partial_\nu
A_\mu-ig[V_\mu,A_\nu]-ig[A_\mu,V_\nu],\nonumber\\
V_\mu&=&\left(\frac{{\bm\tau}}{2}\cdot{\bm\rho}_\mu\right)\mbox{,
} A_\mu=\left(\frac{{\bm\tau}}{2}\cdot{\bm A }_\mu\right)\mbox{, }
\xi=\exp
i\frac{{\bm\tau}\cdot{\bm\pi}}{2f^{(0)}_\pi},\label{not}\end{eqnarray}
where ${\bm\rho}_\mu$, ${\bm\pi}$ are the vector meson $\rho$  and
pseudoscalar pion fields, respectively, ${\bm A}_\mu$ is the axial
vector field [not $a_1$ meson, see Eq.~(\ref{physa1})],
${\bm\tau}$ is the isospin Pauli matrices. Free parameters
$(a_0,b_0,c_0,d_0)$, and $f^{(0)}_\pi$ of the GHLS lagrangian with
index $0$ are bare parameters before renormalization (see below);
$[,]$ stands for commutator. Hereafter the boldface characters,
cross ($\times$), and dot ($\cdot$) stand for vectors, vector
product, and scalar product, respectively, in the isotopic space.

GHLS lagrangian includes also electroweak sector. In what follows
we will neglect the terms quadratic in electroweak coupling
constants keeping only the terms linear in above couplings. These
terms describe the interaction of $\pi$, $\rho$, and $a_1$ mesons
with electroweak gauge bosons and look as \cite{bando88a,bando88}
\begin{eqnarray}
\Delta{\cal L}_{\rm EW}&=&2f_\pi^{(0)2}\bar{g}{\rm
Tr}\left\{a_0\left(\frac{\partial_\mu\xi^\dag\xi+\partial_\mu\xi\xi^\dag}{2i}\frac{\xi^\dag{\cal
L}_\mu\xi+\xi{\cal
R}_\mu\xi^\dag}{2}\right)+\left(d_0+\frac{b_0c_0}{b_0+c_0}\right)\times\right.\nonumber\\&&\left.
A_{(\xi)\mu}\frac{\xi^\dag{\cal L}_\mu\xi-\xi{\cal
R}_\mu\xi^\dag}{2}-a_0gV_\mu\frac{\xi^\dag{\cal L}_\mu\xi+\xi{\cal
R}_\mu\xi^\dag}{2}+b_0ga_\mu\frac{\xi^\dag{\cal L}_\mu\xi-\xi{\cal
R}_\mu\xi^\dag}{2}\right\}.\label{LEW}
\end{eqnarray}
Upon neglecting the weak neutral current contribution,  the
charged weak and electromagnetic sectors are taken into account
via  \cite{bando88a,bando88},
\begin{eqnarray}
\bar{g}{\cal L}_\mu&=&\frac{g_2}{\sqrt{2}}(W^+_\mu T_-+W^-_\mu
T_+)+eQ{\cal A}_\mu\mbox{, } \bar{g}{\cal R}_\mu=eQ{\cal A}_\mu,
\label{LR}\end{eqnarray} $W^\pm_\mu$ are the fields of $W^\pm$
bosons,  $g_2$ is the electroweak $SU(2)$ gauge coupling constant.
In the $SU(2)$ subgroup of the flavor $SU(3)$ group of strong
interactions one has
$T^+=\left(%
\begin{array}{cc}
  0 & V_{ud} \\
  0 & 0 \\
\end{array}%
\right)$, $V_{ud}=\cos\theta_C$ is the element of
Cabibbo-Kobayashi-Maskawa matrix, ${\cal A}_\mu$ stands for the
field of the photon,  $e$ is the elementary charge, and
$Q=\frac{1}{3}\left(%
\begin{array}{cc}
  2 & 0 \\
  0 & -1 \\
\end{array}%
\right)$ is the charge matrix restricted to the sector of
nonstrange mesons. In the spirit of chiral perturbation theory, as
the first step in obtaining necessary terms, one should expand the
matrix $\xi$ into the series over ${\bm\pi}/f_\pi^{(0)}$. The
second step is the renormalization necessary for canonical
normalization of the pion kinetic term. The renormalization is
\cite{bando88a,bando88}
\begin{eqnarray}
f^{(0)}_\pi&=&Z^{-1/2}f_\pi, {\bm\pi}\to Z^{-1/2}{\bm\pi},
(a_0,b_0,c_0,d_0)=Z\times(a,b,c,d),\label{renorm}\end{eqnarray}
where $\left(d_0+\frac{b_0c_0}{b_0+c_0}\right)Z^{-1}=1.$ Close
examination of Eq.~(\ref{LEW}) shows that the expansion includes
the point-like interaction
$\left(\frac{a}{2}-d-\frac{bc}{b+c}\right)W^-_\mu
[{\bm\pi}\times\partial_\mu{\bm\pi}]_{1+i2}.$ Analogous term
appears when one restores electromagnetic field. Since there are
no experimental indications on point-like $\gamma\to\pi^+\pi^-$
vertex, we set
\begin{equation}
\frac{a}{2}-d-\frac{bc}{b+c}=0.\label{nopoint}\end{equation} This
relation removes also the above point-like $W^-\to\pi^-\pi^0$
vertex.

{\bf 3. The amplitude of the transition ${\bm
W^-\to2\pi^-\pi^+}$.} Expanding the GHLS lagrangian into the
series in the ratio of the pion momentum to the pion decay
constant $f_\pi=92.4$ MeV, first, one obtains the relations
\begin{equation}
g_{\rho\pi\pi}=\frac{ag}{2}\mbox{, } m^2_\rho=ag^2f^2_\pi\mbox{, }
m^2_{a_1}=(b+c)g^2f^2_\pi.\label{masscoup}\end{equation} We fix
hereafter $g_{\rho\pi\pi}$ from the experimental value of the
$\rho^0\to\pi^+\pi^-$ decay width leaving  $a$ as free parameter.
Second, the lagrangian describing the decay $a_1\to3\pi$ is found
to be
\begin{eqnarray}
{\cal L}_{a_13\pi}&=&-\frac{r}{f_\pi}(\partial_\mu{\bm
a}_\nu-\partial_\nu{\bm
a}_\mu)[{\bm\rho}_\mu\times\partial_\nu{\bm\pi}]+
\frac{\alpha_5}{f_\pi}{\bm
a}_\mu[(\partial_\mu{\bm\rho}_\nu-\partial_\nu{\bm\rho}_\mu)\times\partial_\nu{\bm\pi}]-\nonumber\\&&
\frac{r^2}{gf^3_\pi}(\alpha_5-r)[{\bm
a}_\mu\times\partial_\nu{\bm\pi}]\cdot[\partial_\mu{\bm\pi}\times\partial_\nu{\bm\pi}]-
\frac{r}{2gf^3_\pi}\partial_\mu{\bm
a}_\nu\cdot[{\bm\pi}\times[\partial_\mu{\bm\pi}\times\partial_\nu{\bm\pi}]].\label{La13pi}
\end{eqnarray}
The amplitude of the decay
$a_1^-(q)\to\pi^+(q_1)\pi^-(q_2)\pi^-(q_3)$ calculated from
Eq.~(\ref{La13pi}) can be written as follows:
$M[a_1^-(q)\to\pi^+(q_1)\pi^-(q_2)\pi^-(q_3)]\equiv M_{a_13\pi}$,
\begin{eqnarray}
iM_{a_13\pi}&=&\frac{agr}{2f_\pi}
\epsilon_\mu\left(A_1q_{1\mu}+A_2q_{2\mu}+A_3q_{3\mu}\right),
\label{ma13pi}\end{eqnarray} where $\epsilon_\mu$ is the
polarization four-vector of $a_1$ meson, and
$A_1=(1+\hat{P}_{23})\tilde{A_1}$, where
\begin{eqnarray*}
\tilde{A_1}&=&\frac{\beta[(q_3,q_1-q_2)-(q,q_3)+m^2_\pi]-(q,q_3)}{D_\rho(q_1+q_2)}
+\frac{4r^2(\beta-1)(q_2,q_3)+(q,q)-(q,q_1)}{2m^2_\rho},\nonumber\\
A_2&=&\frac{\beta[(q_3,q_1-q_2)+(q,q_3)-m^2_\pi]+(q,q_3)}{D_\rho(q_1+q_2)}+\frac{(q_2,q_1-q_3)}{D_\rho(q_1+q_3)}
-\frac{2r^2(\beta-1)(q_1,q_3)+(q,q_1)}{m^2_\rho}. \label{A12}
\end{eqnarray*}
Hereafter $\hat{P}_{ij}$ interchanges pion momenta $q_i$ and
$q_j$, $(q_i,q_j)$ stands for the Lorentz scalar product of
four-vectors, and $A_3=\hat{P}_{23}A_2$. Parameters $r$ and
$\beta$ are the combinations of the GHLS parameters:
\begin{eqnarray}
r&=&\frac{b}{b+c}\mbox{,
}\beta=\frac{\alpha_5}{r}.\label{rbet}\end{eqnarray} Notice that
the amplitude (\ref{ma13pi}) respects the Adler condition: it
vanishes in the chiral limit $m^2_\pi\to 0$ when the four-momentum
of any final pion  vanishes. Such a property is the manifestation
of the chiral invariance.

\begin{figure}
\begin{center}
\includegraphics[width=80mm]{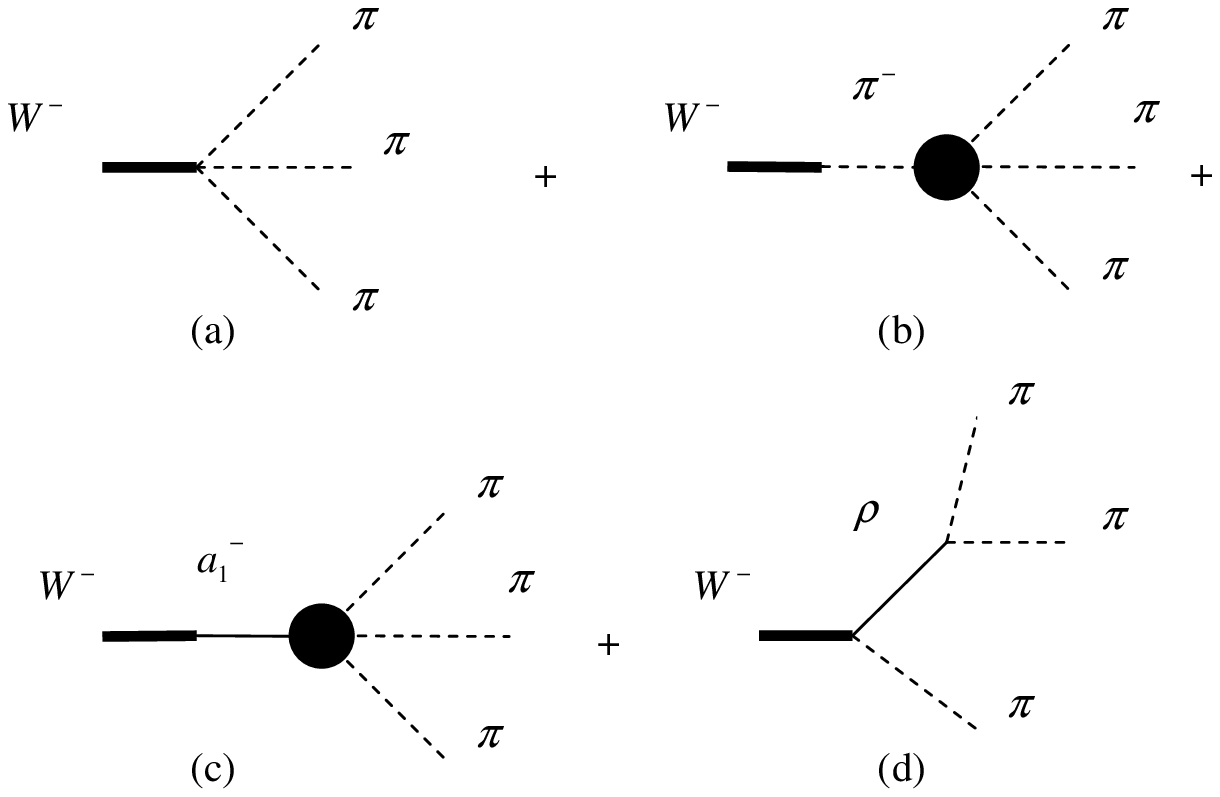}
\caption{\label{taucdiag}Diagrams schematically describing the
transition $W^-\to\pi^-\pi^-\pi^+$. Shaded circles depict the
transition including both  the point-like and $\rho$-exchange
contributions. Permutations of pion momenta are understood.}
\end{center}
\end{figure}

The amplitude of the decay $\tau^-\to\pi^-\pi^-\pi^+\nu_\tau$
incorporates the transition $W^-\to\pi^-\pi^-\pi^+$. In GHLS, the
latter is given by the diagrams shown in Fig.~\ref{taucdiag}.
Necessary terms are obtained from the low momentum expansion of
electroweak piece of GHLS lagrangian Eq.~(\ref{LEW}) and look like
\begin{eqnarray}
\Delta{\cal L}_{\rm EW}&=&\frac{1}{2}g_2V_{ud}{\bm
W}_{\mu\bot}\left(-f_\pi\partial_\mu{\bm\pi}_\bot+\frac{1}{3f_\pi}
[{\bm\pi}\times[{\bm\pi}\times\partial_\mu{\bm\pi}]]_\bot
+bgf^2_\pi{\bm
a}_{\mu\bot}+agf_\pi[{\bm\pi}\times{\bm\rho}_\mu]_\bot\right)\nonumber\\&&
-eg{\cal A}_\mu\left\{a\rho^0_\mu(f^2_\pi-\pi^+\pi^-)-
\frac{2\pi^+\pi^-}{3gf^2_\pi}[{\bm\pi}\times\partial_\mu{\bm\pi}]_3
\left(\frac{7}{8}a-rc\right)+ bf_\pi[{\bm\pi}\times{\bm
a}_\mu]_3\right\}, \label{DeltaL}
\end{eqnarray}
where  the vector ${\bm V}_\bot=(V_1,V_2)$ denotes transverse
charged components of the isotopic vector. The amplitude of the
decay $W^-(q)\to\pi^+(q_1)\pi^-(q_2)\pi^-(q_3)$ corresponding to
the diagrams Fig.~\ref{taucdiag} is
$iM=\frac{g_2V_{ud}}{2f_\pi}\epsilon_\mu^{(W)}J_\mu$, where
$\epsilon_\mu^{(W)}$ is the polarization four-vector of $W^-$
boson and the axial decay current $J_\mu$ looks like
\begin{eqnarray}
J_\mu&=&-q_{1\mu}+\frac{q_\mu}{D_\pi(q)}\left[m^2_\pi-(q,q_1)+\frac{am^2_\rho}{2}(1+\hat{P}_{23})
\frac{(q_2,q_1-q_3)}{D_\rho(q_1+q_3)}\right]-\frac{ar^2m^2_{a_1}}{2D_{a_1}(q)}\times\nonumber\\&&
\left\{A_1q_{1\mu}+A_2q_{2\mu}+A_3q_{3\mu}-\frac{2q_\mu}{m^2_{a_1}}(1+\hat{P}_{23})
\left[(m^2_\pi+(q_1,q_2))(q_3,q_1-q_2)\times\right.\right.\nonumber\\&&\left.\left.
\left(\frac{\beta}{D_\rho(q_1+q_2)}-\frac{r^2(\beta-1)}{m^2_\rho}\right)\right]\right\}+\frac{am^2_\rho}{2}
(1+\hat{P}_{23})\frac{(q_1-q_3)_\mu}{D_\rho(q_1+q_3)}.
\label{Jdec}
\end{eqnarray}
In the above expressions, $D_\rho$, $D_\pi$, and $D_{a_1}$ are the
inverse propagators of $\pi$, $\rho$, and $a_1$ mesons,
respectively. The terms corresponding to the diagrams (a), (b),
(c), and (d) in Fig.~\ref{taucdiag} are easily identified by these
propagators.

The spectrum of the three pion state in the decay
$\tau^-\to\pi^+\pi^-\pi^-\nu_\tau$ normalized to its branching
fraction is
\begin{eqnarray}
\frac{dB}{ds}&=&\frac{(G_FV_{ud})^2(m^2_\tau-s)^2}{2\pi(2m_\tau)^3\Gamma_\tau}
\left[(m^2_\tau+2s)\rho_t(s)+m^2_\tau\rho_l(s)\right],\end{eqnarray}
$s=q^2$, $G_F$ is the Fermi constant, and $\Gamma_\tau$ is the
width of $\tau$ lepton. The transverse and longitudinal spectral
functions are, respectively,
\begin{eqnarray}
\rho_t(s)&=&\frac{1}{3\pi sf^2_\pi}\int
d\Phi_{3\pi}\left[\frac{|(q,J)|^2}{s}-(J,J^\ast)\right]\mbox{, }
\rho_l(s)=\frac{1}{\pi s^2f^2_\pi}\int d\Phi_{3\pi}|(q,J)|^2,
\end{eqnarray}
where $d\Phi_{3\pi}$ is the element of Lorentz-invariant phase
space  volume of the system $\pi^-\pi^-\pi^+$. The numerical
integration shows that $\rho_l$ is by about three orders of
magnitude smaller than $\rho_t$ in all allowed kinematical range
$9m^2_\pi<s<m^2_\tau$ and hence can be neglected.

{\bf 4. Results for $\tau^-\to\pi^+\pi^-\pi^-\nu_\tau$.} The
"canonical" choice of free GHLS parameters \cite{bando88}
\begin{equation}
(a,b,c,d,\alpha_4,\alpha_5,\alpha_6)=
(2,2,2,0,-1,1,1),\label{canon}\end{equation}and
$\alpha_1=\alpha_2=\alpha_3=0$, results in the spectrum which
disagrees with the data both in lower branching ratio
$B_{\tau^-\to\pi^+\pi^-\pi^-\nu_\tau} \approx6\%$ and in the shape
of the spectrum.  Upon the variation of  free parameters of the
single $a_1$ resonance contribution listed in Eq.~(\ref{canon})
one obtains  the can  reproduce the branching ratio
$B_{\tau^-\to\pi^+\pi^-\pi^-\nu_\tau} \approx9\%$ but the shape of
the spectrum is not reproduced. Inclusion of additional higher
derivative terms \cite{li} to the suggested in
Refs.~\cite{bando88a,bando88} and subjected to the fitting in the
present work minimal set Eq.~(\ref{canon}) cannot improve the
situation. Indeed, even the minimal set Eq.~(\ref{canon}) results
in a rather fast growth of the $a_1\to3\pi$ decay width with the
energy increase. Additional higher derivative terms  would make
the growth to be explosive. Restricting such a growth would
require phenomenological form factors with free parameters. We
believe that the dynamical explanation of the shape of the
spectrum  based on additional axial vector resonances
$a_1^\prime$, $a_1^{\prime\prime}$ would be preferable. Note that
there are indications on such resonances, both theoretical
\cite{isgur87,isgur85} and experimental \cite{pdg,amelin,cleo}.

\begin{figure}
\begin{center}
\includegraphics[width=80mm]{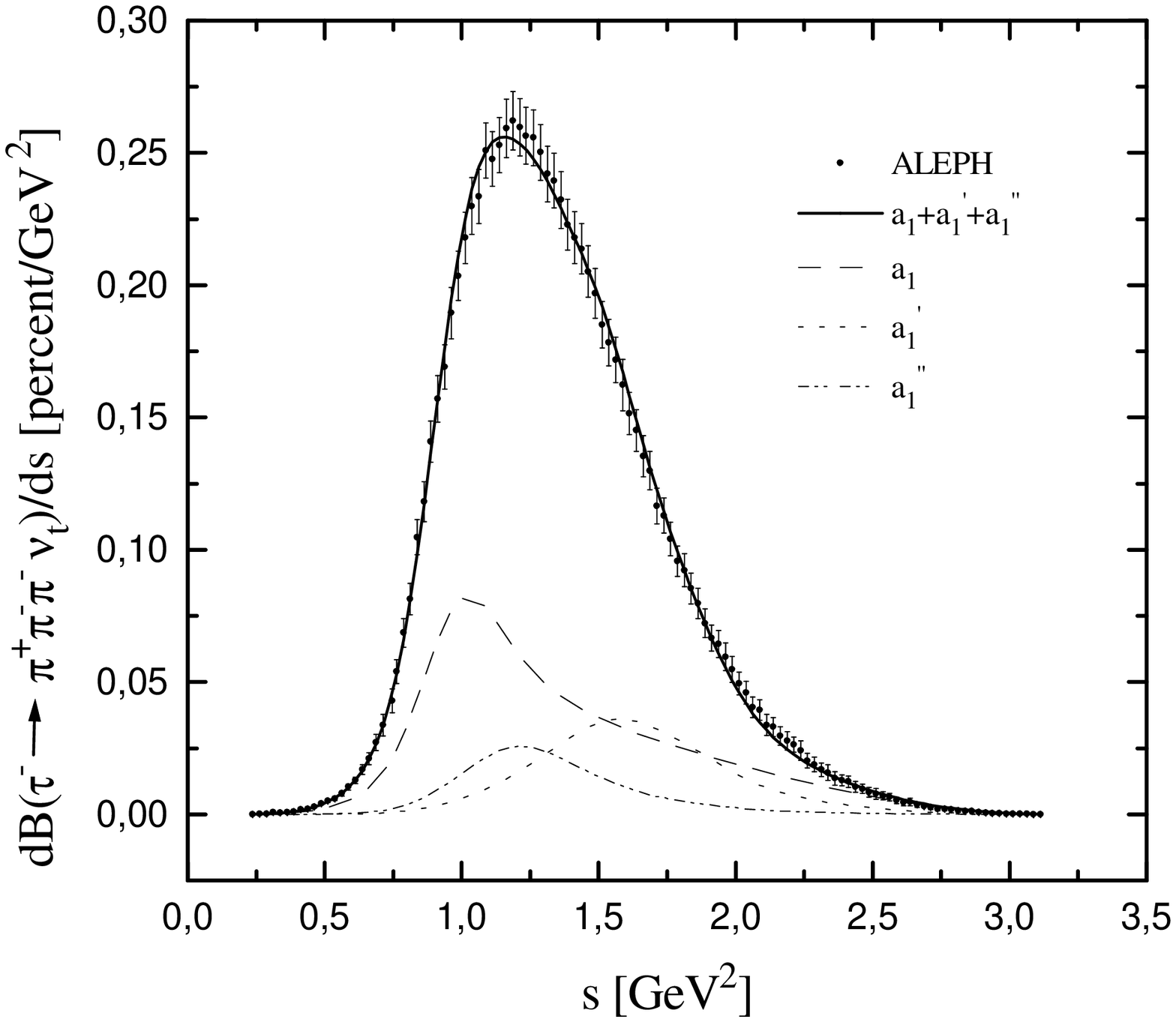}
\caption{\label{spec_const}The spectrum of $\pi^+\pi^-\pi^-$ in
$\tau$ decay in the variant A of the Table \ref{fitresults}. The
dashed line corresponds to the sum of the diagrams
Fig.~\ref{taucdiag}. See the text for more  detail.}
\end{center}
\end{figure}

      Taking $a_1^\prime$, $a_1^{\prime\prime}$ into account  reduces to adding
two diagrams similar to one in Fig.~\ref{taucdiag}(c), with the
replacement of $a_1(1260)$ by $a^\prime_1$ and
$a^{\prime\prime}_1$. Since there is no available information
concerning their couplings, the above resonances  are included in
a way analogous to $a_1(1260)$. This prescription results in the
amplitudes of the decays $a^\prime_1,a^{\prime\prime}_1\to3\pi$
vanishing when the four-momentum of any final pion vanishes. That
is,  the way of inclusion additional resonances respects chiral
symmetry. The total set of the fitted parameters is first taken to
be
$$(m_{a_1},a,r,\beta,m_{a^\prime_1},a^\prime,r^\prime,\beta^\prime,w^\prime,
m_{a^{\prime\prime}_1},a^{\prime\prime},r^{\prime\prime},\beta^{\prime\prime},w^{\prime\prime}).$$
The parameters $a^\prime$, $r^\prime$, $\beta^\prime$ characterize
the $a^\prime_1\to3\pi$ decay amplitude similar to
Eq.~(\ref{ma13pi}), (\ref{A12}) in the case of $a_1(1260)\to3\pi$,
while $w^\prime$  parameterizes  the coupling $a^\prime_1\rho\pi$
as $g_{\rho\pi\pi}w^\prime r^\prime/f_\pi$. Compare with
Eq.~(\ref{ma13pi}). Analogously for $a_1^{\prime\prime}$.  The fit
chooses $w^\prime=1$ and turns out to be insensitive to this
parameter leaving $\chi^2/N_{\rm d.o.f}=122/102$. The quality of
the fit can be considerably improved upon fixing $w^\prime=1$ but
adding new parameter $\psi^\prime$-the phase of the $a^\prime_1$
contribution. Such phase imitates possible mixing among $a_1$,
$a_1^\prime$, $a_1^{\prime\prime}$ resonances. The results of such
type of the fit are given in the column variant A of the Table
\ref{fitresults}.
\begin{table}
\caption{\label{fitresults}The values of free parameters of GHLS
model obtained from the unconstrained fit of the ALEPH data  on
the decay $\tau^-\to\pi^+\pi^-\pi^-\nu_\tau$ \cite{aleph05}
(variant A), and the fit with the constrain $a=2$ preserving
universality (variant B). Also shown are the corresponding
calculated original \cite{bando88a,bando88} GHLS parameters and
the   magnitudes of branching fractions of the above decay.  }
\begin{center}
\begin{tabular}{|c|c|c|}
\hline
parameter&variant A&variant B\\
\hline
$m_{a_1}$[GeV]&$1.332\pm0.015$&$1.139\pm0.016$\\
$a$&$1.665\pm0.011$&$\equiv 2$\\
$b$(calculated)&$1.35\pm0.05$&$0.52\pm0.03$\\
$c$(calculated)&$2.72\pm0.08$&$3.74\pm0.11$\\
$d$(calculated)&$-0.07\pm0.03$&$0.54\pm0.03$\\
$\alpha_4$(calculated)&$-10\pm1$&$-27\pm2$\\
$\alpha_5$(calculated)&$2.82\pm0.06$&$1.94\pm0.15$\\
$r$&$0.332\pm0.007$&$0.122\pm0.006$\\
$\beta$&$8.5\pm0.3$&$15.9\pm0.9$\\
$m_{a^\prime_1}$[GeV]&$1.59\pm0.01$&$1.76\pm0.01$\\
$a^\prime$&$0.99\pm0.01$&$1.09\pm0.01$\\
$r^\prime$&$0.96\pm0.01$&$0.90\pm0.01$\\
$\beta^\prime$&$0.07\pm0.02$&$0.28\pm0.02$\\
$w^\prime$&$\equiv1$&$\equiv1$\\
$\psi^\prime$&$28^\circ\pm1^\circ$&$48^\circ\pm1^\circ$\\
$m_{a^{\prime\prime}_1}$[GeV]&$1.88\pm0.02$&$2.27\pm0.02$\\
$a^{\prime\prime}$&$0.46\pm0.01$&$0.59\pm0.01$\\
$r^{\prime\prime}$&$1.45\pm0.02$&$1.56\pm0.02$\\
$\beta^{\prime\prime}$&$0.91\pm0.05$&$0.91\pm0.03$\\
$w^{\prime\prime}$&$1.14\pm0.01$&$1.37\pm0.01$\\
$\psi^{\prime\prime}$&$\equiv0^\circ$&$\equiv0^\circ$\\
$B_{\tau^-\to\pi^+\pi^-\pi^-\nu_\tau}$&$(9.05\pm0.16)\%$&$(9.00\pm0.15)\%$\\
$\chi^2/N_{\rm d.o.f}$&79/102&70/103\\
\hline
\end{tabular}
\end{center}
\end{table}
The corresponding  spectrum is shown in Fig.~\ref{spec_const}.
Using Eq.~(\ref{masscoup}), (\ref{rbet}), and obtaining
$g_{\rho\pi\pi}=5.95$ from $\Gamma_{\rho\pi\pi}$ \cite{pdg} one
can compare the fitted GHLS parameters with the "canonical" ones
Eq.~(\ref{canon}). To this end one should use Eq.~(\ref{nopoint})
and (\ref{masscoup}) to obtain
\begin{eqnarray}
b&=&r\left(\frac{m_{a_1}a}{2f_\pi g_{\rho\pi\pi}}\right)^2\mbox{,
} c=(1-r)\left(\frac{m_{a_1}a}{2f_\pi
g_{\rho\pi\pi}}\right)^2\mbox{, }
d=\frac{a}{2}-r(1-r)\left(\frac{m_{a_1}a}{2f_\pi g_{\rho\pi\pi}}\right)^2,\nonumber\\
\alpha_4&=&1-2\beta(1-r)\mbox{, } \alpha_5=\beta r\mbox{, }
\alpha_6=\alpha_5. \label{realparam}\end{eqnarray} These GHLS
parameters are marked in the Table \ref{fitresults} as
"calculated". Since the basis of inclusion of  heavier resonances
$a_1^\prime$ and $a_1^{\prime\prime}$ here is  purely
phenomenological, specifically, there is no analog of gauge
coupling constant $g$, we do not recalculate
$(a^\prime,b^\prime,c^\prime,d^\prime,\cdots)$ and
$(a^{\prime\prime},b^{\prime\prime},c^{\prime\prime},d^{\prime\prime},\cdots)$
similar to Eq.~(\ref{realparam}). One can see that the obtained
$a=1.665\pm0.011$ is in disagreement with the universality
condition $g_{\rho\pi\pi}=g$, which demands $a=2$, see
Eq.~(\ref{masscoup}). Hence we  fulfill also the partially
constrained fit with $a\equiv2$, in order to preserve universality
of the $\rho$ couplings. The results are presented as the variant
B in the Table \ref{fitresults}. The total spectrum in this
variant is not shown because it looks the same as in
Fig.~\ref{spec_const}. Note that in the variant A  the visible
$a^{\prime\prime}_1$ peak position is lower than that of
$a^\prime_1$ despite of the fact that their bare masses are in
opposite relation, see the Table \ref{fitresults}. This  can be
explained as follows. The dominant decay mode of $a_1^\prime$,
$a_1^{\prime\prime}$ resonances  is the $3\pi$ one. Its partial
width grows rapidly with energy increase reaching the figures
compatible with bare mass itself. The combined action of the
strong energy dependence of the partial width and its large
magnitude shifts the visible peak towards the lower energies
\cite{ach00a}. The evaluation shows that the width of
$a_1^{\prime\prime}$ and its growth are stronger as compared to
$a_1^\prime$. Hence the visible position of the former appears at
lower energy than the visible position of $a_1^\prime$.

\begin{figure}
\begin{center}
\includegraphics[width=80mm]{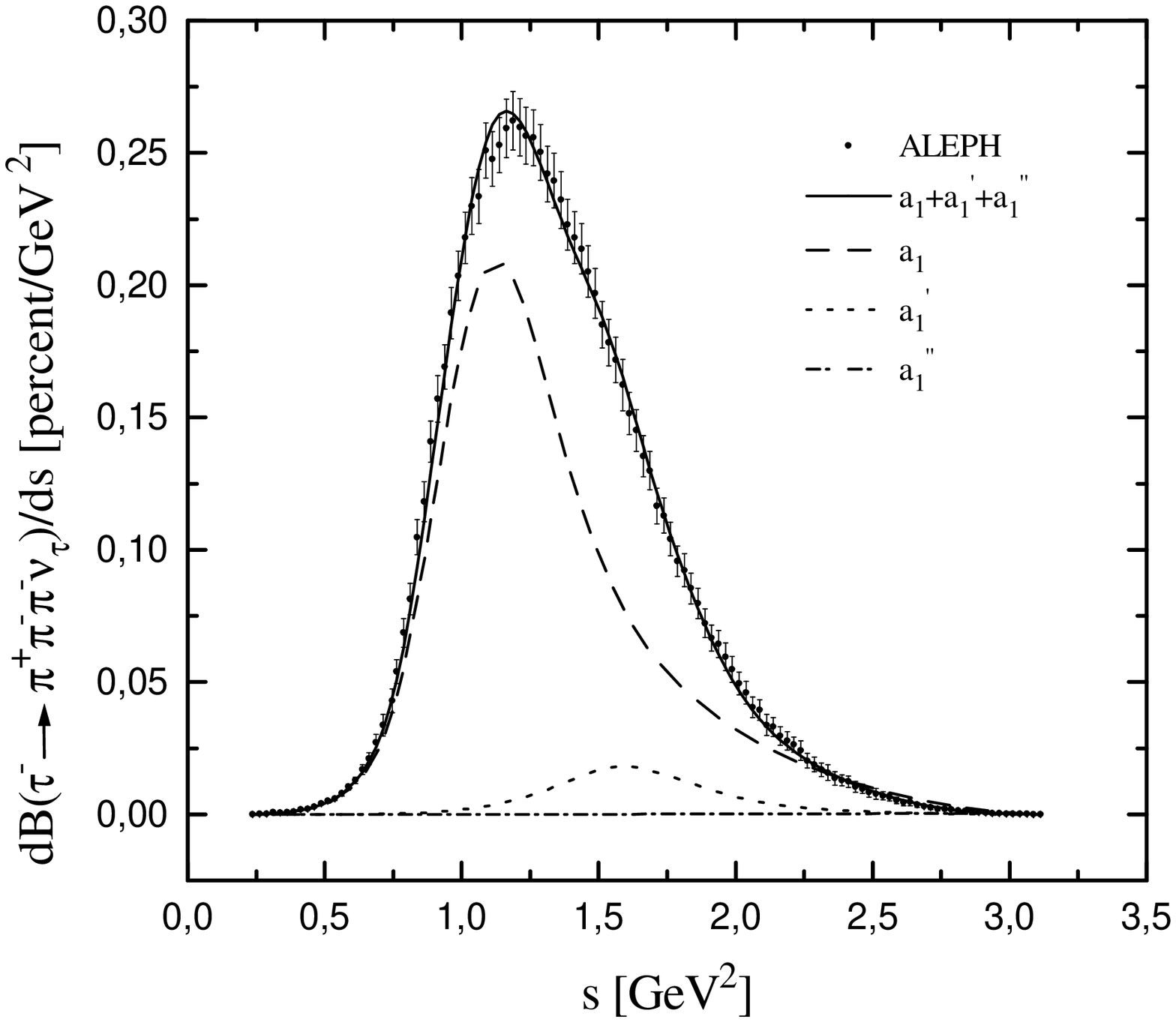}
\caption{\label{fixr_const}The same as in Fig.~\ref{spec_const},
but for the parameters of the variant C of the Table
\ref{fitresCD}.}
\end{center}
\end{figure}

Of a special interest is the width of the radiative decay
$a_1^\pm\to\pi^\pm\gamma$.    This decay originates  from both the
$a_1\to\rho\pi$ transition followed by the transition
$\rho\to\gamma$ and by the direct $a_1\to\pi\gamma$ transition.
The necessary amplitudes can be read off Eq.~(\ref{DeltaL}). The
resulting $a_1^\pm\to\pi^\pm\gamma$ decay width is
\begin{eqnarray}
\Gamma_{a^\pm_1\to\pi^\pm\gamma}&=&\frac{\alpha
am^3_{a_1}}{24m^2_\rho}\left[r(\beta-1)\right]^2
\left(1-\frac{m^2_\pi}{m^2_{a_1}}\right)^3,\label{Gammaa1pigamma}\end{eqnarray}
where $\alpha$ is the fine structure constant. Notice, that the
above expression for $\Gamma_{a^\pm_1\to\pi^\pm\gamma}$ is written
with the counter terms taken into account. The
$a^\pm_1\to\pi^\pm\gamma$ decay amplitude without counter terms is
proportional to the combination  $b-arm^2_{a_1}/m^2_\rho$ which
vanishes at any choice of GHLS parameters because of the relations
(\ref{masscoup}) and (\ref{rbet}).  The evaluation of
$\Gamma_{a^\pm_1\to\pi^\pm\gamma}$ with the parameters from the
variants A and B of the Table \ref{fitresults} gives the figures
of the order of few MeV due to large values of $\beta$  in the
Table \ref{fitresults} chosen by the fits. This is in disagreement
with the  measured $\Gamma_{a^\pm_1\to\pi^\pm\gamma}=640\pm246$
keV \cite{ziel}.  Hence, one should further constrain the fit in
order to incorporate the above radiative width upon expressing
parameter $\beta$ from Eq.~(\ref{Gammaa1pigamma}). When fitting,
the central value of the $a_1(1260)$ radiative width. To provide
the universality of the $\rho$ couplings, $a=2$ is kept fixed,
too. It is found out that the fit with the fixed parameters $a$
and $\beta$  gives rather poor description with $\chi^2/N_{\rm
d.o.f}=209/102$. The peculiar feature of the fit is that it
chooses $\psi^\prime\approx0$, the phase of the $a_1^\prime$
contribution, but $\chi^2$ is almost insensitive to the rather
wide  variations around above central value. Hence, we fix
$\psi^\prime\equiv0$, but introduce a new free parameter $\gamma$
whose meaning is $\gamma=m_{a_1}\Delta\Gamma_{a_1}$, where
$\Delta\Gamma_{a_1}$ effectively takes into account the
contributions to the $a_1$ resonance width other than
$\rho\pi+3\pi\to3\pi$ one, for example,
$a_1\to\rho^\prime\pi\to3\pi$, $K\bar K\pi$. They may be effective
for the off-mass-shell $a_1$ meson.  The results of such type of
the fit   are presented as the variant C in the Table
\ref{fitresCD}.
\begin{table}
\caption{\label{fitresCD}The values of free parameters of GHLS
model obtained from the  fit of the ALEPH data  on the decay
$\tau^-\to\pi^+\pi^-\pi^-\nu_\tau$ \cite{aleph05} constrained in a
way as to fix $a\equiv2$ and $\Gamma_{a_1^\pm\to\pi^\pm\gamma}$.
Variant C is  the fit including
$a_1+a_1^\prime+a_1^{\prime\prime}$ contributions. Variant D
includes only $a_1+a_1^\prime$ ones. Also shown are the
corresponding calculated original \cite{bando88a,bando88} GHLS
parameters and the   magnitudes of branching fractions of the
above decay.  }
\begin{center}
\begin{tabular}{|c|c|c|}
\hline
parameter&variant C&variant D\\
\hline
$m_{a_1}$[GeV]&$1.368\pm0.006$&$1.401\pm0.006$\\
$a$&$\equiv2$&$\equiv 2$\\
$b$(calculated)&$4.89\pm0.07$&$5.37\pm0.06$\\
$c$(calculated)&$1.30\pm0.07$&$1.12\pm0.05$\\
$d$(calculated)&$-0.03\pm0.06$&$0.07\pm0.04$\\
$\alpha_4$(calculated)&$0.66\pm0.06$&$0.45\pm0.05$\\
$\alpha_5$(calculated)&$1.29\pm0.10$&$1.31\pm0.10$\\
$r$&$0.790\pm0.008$&$0.827\pm0.006$\\
$\beta$(calculated)&$1.63\pm0.12$&$1.58\pm0.12$\\
$\gamma$[GeV$^2$]&$0.31\pm0.01$&$0.35\pm0.02$\\
$m_{a^\prime_1}$[GeV]&$1.422\pm0.007$&$1.513\pm0.001$\\
$a^\prime$&$1.80\pm0.03$&$2.01\pm0.03$\\
$r^\prime$&$0.386\pm0.005$&$0.370\pm0.006$\\
$\beta^\prime$&$0.96\pm0.05$&$0.82\pm0.05$\\
$w^\prime$&$1.19\pm0.01$&$1.18\pm0.02$\\
$\psi^\prime$&$\equiv0$&$39^\circ\pm1^\circ$\\
$m_{a^{\prime\prime}_1}$[GeV]&$1.800\pm0.007$&$-$\\
$a^{\prime\prime}$&$-0.32\pm0.02$&$-$\\
$r^{\prime\prime}$&$0.36\pm0.02$&$-$\\
$\beta^{\prime\prime}$&$-0.2\pm0.2$&$-$\\
$w^{\prime\prime}$&$0.30\pm0.04$&$-$\\
$\psi^{\prime\prime}$&$10^\circ\pm8^\circ$&$-$\\
$B_{\tau^-\to\pi^+\pi^-\pi^-\nu_\tau}$&$(8.97\pm0.13)\%$&$(8.96\pm0.17)\%$\\
$\chi^2/N_{\rm d.o.f}$&45/102&95/107\\
\hline
\end{tabular}
\end{center}
\end{table}
The spectrum of the system $\pi^+\pi^-\pi^-$ evaluated with the
parameters of variant C is shown with the solid line in
Fig.~\ref{fixr_const}. Note that the found
$\gamma=m_{a_1}\Delta\Gamma_{a_1}\sim0.3$ GeV$^2$ corresponds to
the portion of the $a_1$ decay channels different from
$\rho\pi+3\pi\to3\pi$ one, at the level
$\Delta\Gamma_{a_1}/\Gamma_{a_1\to3\pi}\sim0.02$. This estimate
can be obtained from the calculated
$\Gamma_{a_1\to\rho\pi+3\pi\to3\pi}$. The above estimate
demonstrates that the additional contribution to the $a_1$ width
beside the GHLS one is very small. Since  the contribution of the
resonance $a_1^{\prime\prime}$ is rather small, see
Fig.~\ref{fixr_const}, we fulfill the fit in which the
contribution of the resonance is absent. The parameters found in
such type of the fit are listed as the variant D in the Table
\ref{fitresCD}.

\begin{figure}[h]
\begin{center}
  \includegraphics[width=8cm]{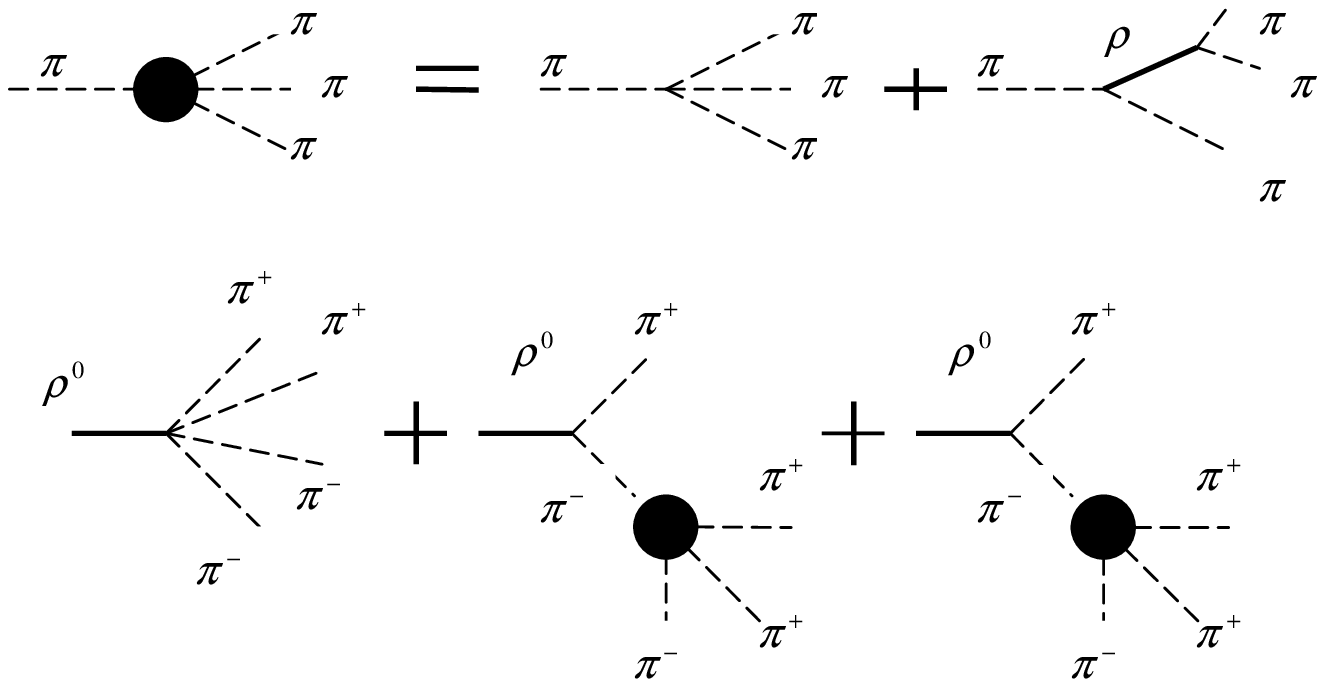}
  \caption{\label{rho4pich}The diagrams due to
HLS lagrangian. Shaded circles in the $\rho\to4\pi$ diagrams stand
for the $\pi\to3\pi$ transition shown in the first line of this
figure.}
\end{center}
 \end{figure}

\begin{figure}[h]
\begin{center}
  \includegraphics[width=8cm]{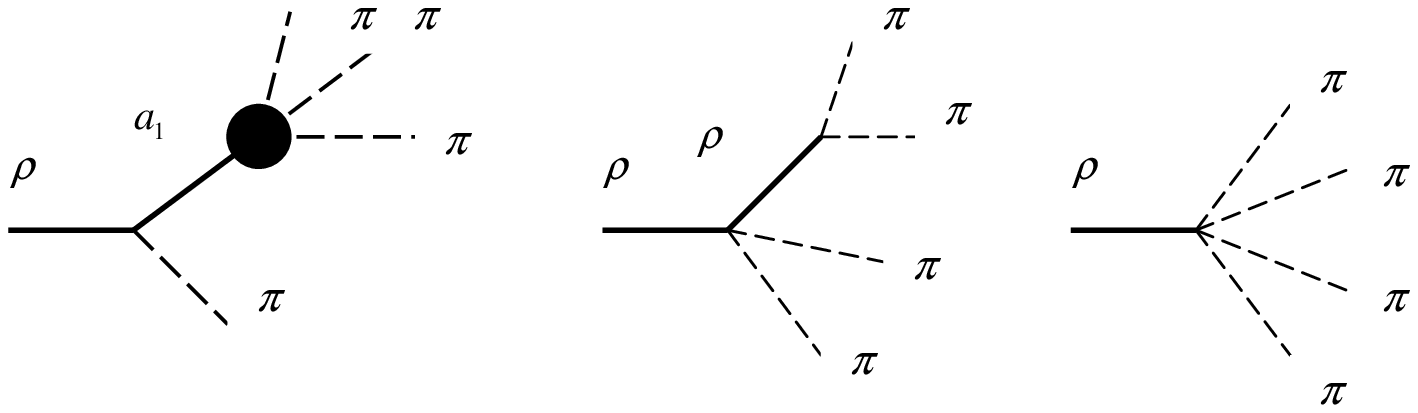}
  \caption{\label{countterm}The diagrams due to
$a_1\rho\pi$ and $\rho\rho\pi\pi$couplings (GHLS). The shaded
circle stands for the total $a_1\to3\pi$ amplitude similar to
Eq.~(\ref{ma13pi}).}
\end{center}
  \end{figure}

\begin{figure}[h]
\begin{center}
  \includegraphics[width=9cm]{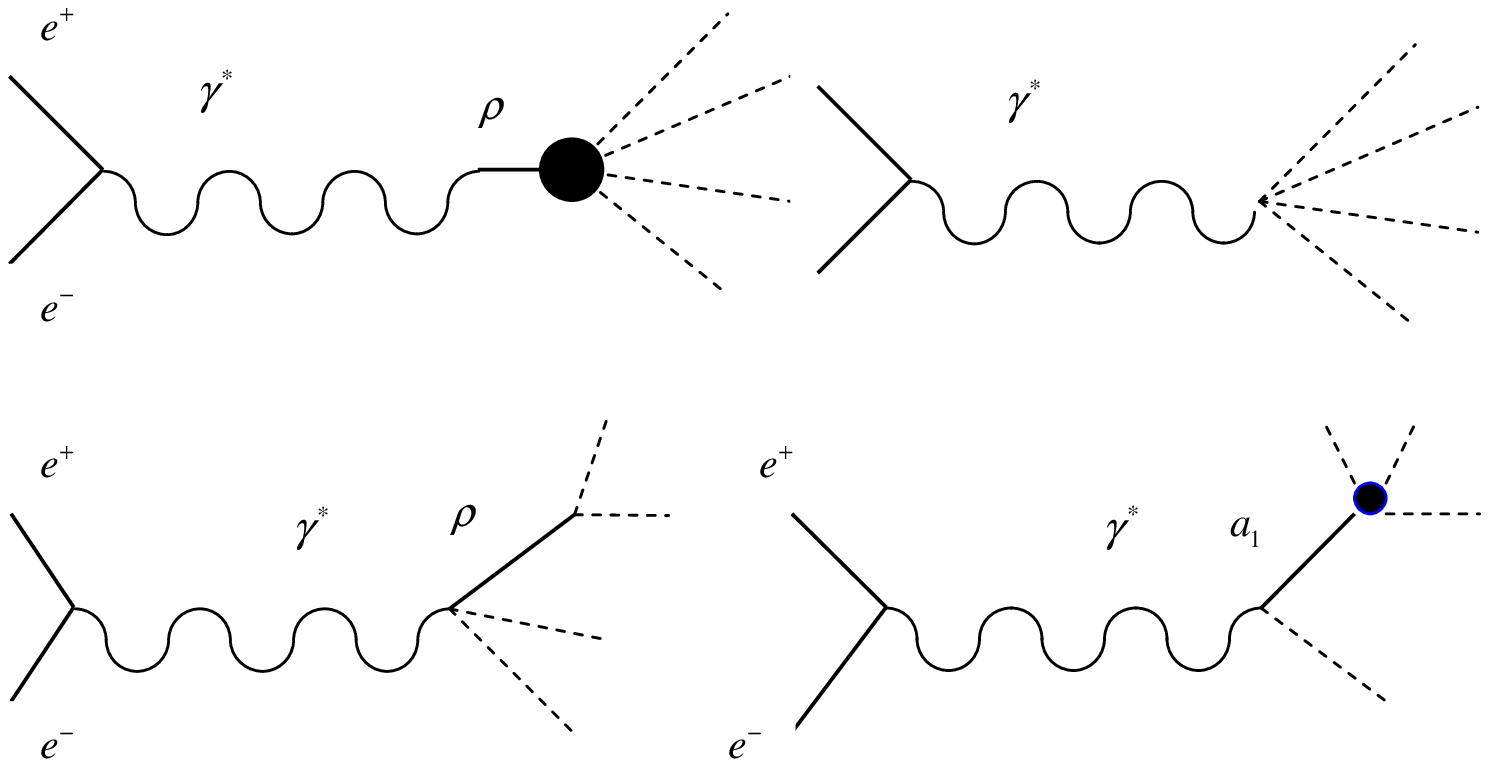}
  \caption{\label{crosssec}Diagrams describing process $e^+e^-\to\pi^+\pi^-\pi^+\pi^-$.
  Shaded circle in the first diagram stands for the sum of the $\rho\to4\pi$
  diagrams in Fig.~\ref{rho4pich} and \ref{countterm}. The last three diagrams are
  due to the point-like $\gamma\to4\pi$, $\gamma\to a_1\pi$, and $\gamma\to\rho\pi\pi$ couplings from the
  lagrangian (\ref{DeltaL}).}
\end{center}
 \end{figure}

{\bf 5. GHLS  and the reaction $e^+e^-\to\pi^+\pi^-\pi^+\pi^-$ at
$\sqrt{s}\leq$ 1 GeV.} Because the evaluation of the  four pion
decay width of the $\rho$-like resonances is very time-consuming,
we study the predictions of GHLS model for the reaction
$e^+e^-\to\pi^+\pi^-\pi^+\pi^-$ with the "canonical" choice of
free parameters (\ref{canon}). The set of the diagrams necessary
for calculation of the amplitude is shown in Fig.~\ref{rho4pich},
\ref{countterm}, and \ref{crosssec}.
This set includes the resonance production $e^+e^-\to
R\to\pi^+\pi^-\pi^+\pi^-$, where
$R=\rho,\rho^\prime,\rho^{\prime\prime}$ (see below), and the
point-like transitions due to the GHLS electromagnetic coupling
(\ref{DeltaL}) where free GHLS parameters are chosen in accord
with (\ref{canon}).

\begin{figure}[h]
\begin{center}
  \includegraphics[width=7cm]{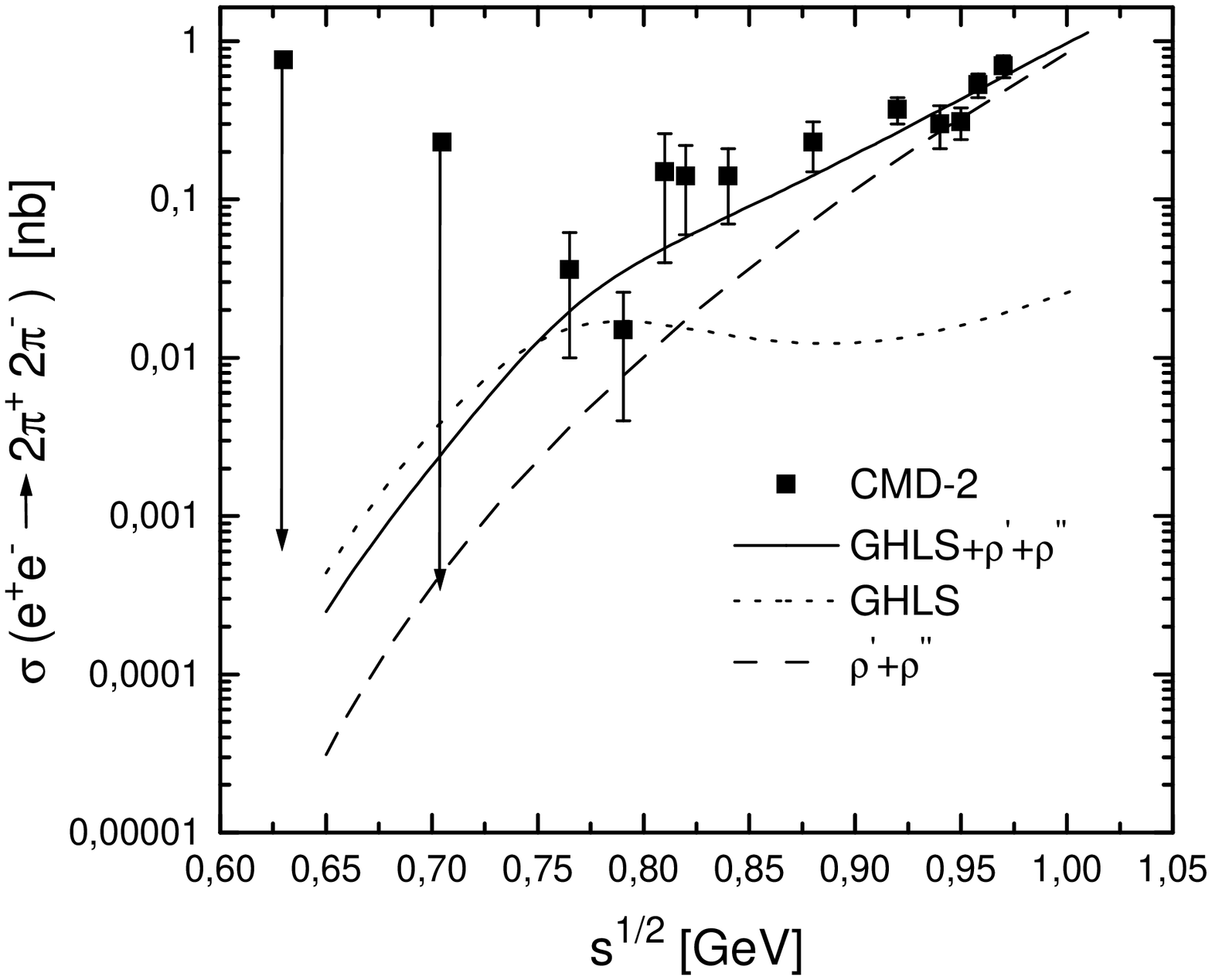}
  \caption{\label{fitcmd2}Fitting CMD-2 data \cite{cmd2}.}
\end{center}
\end{figure}

\begin{figure}[h]
\begin{center}
  \includegraphics[width=7cm]{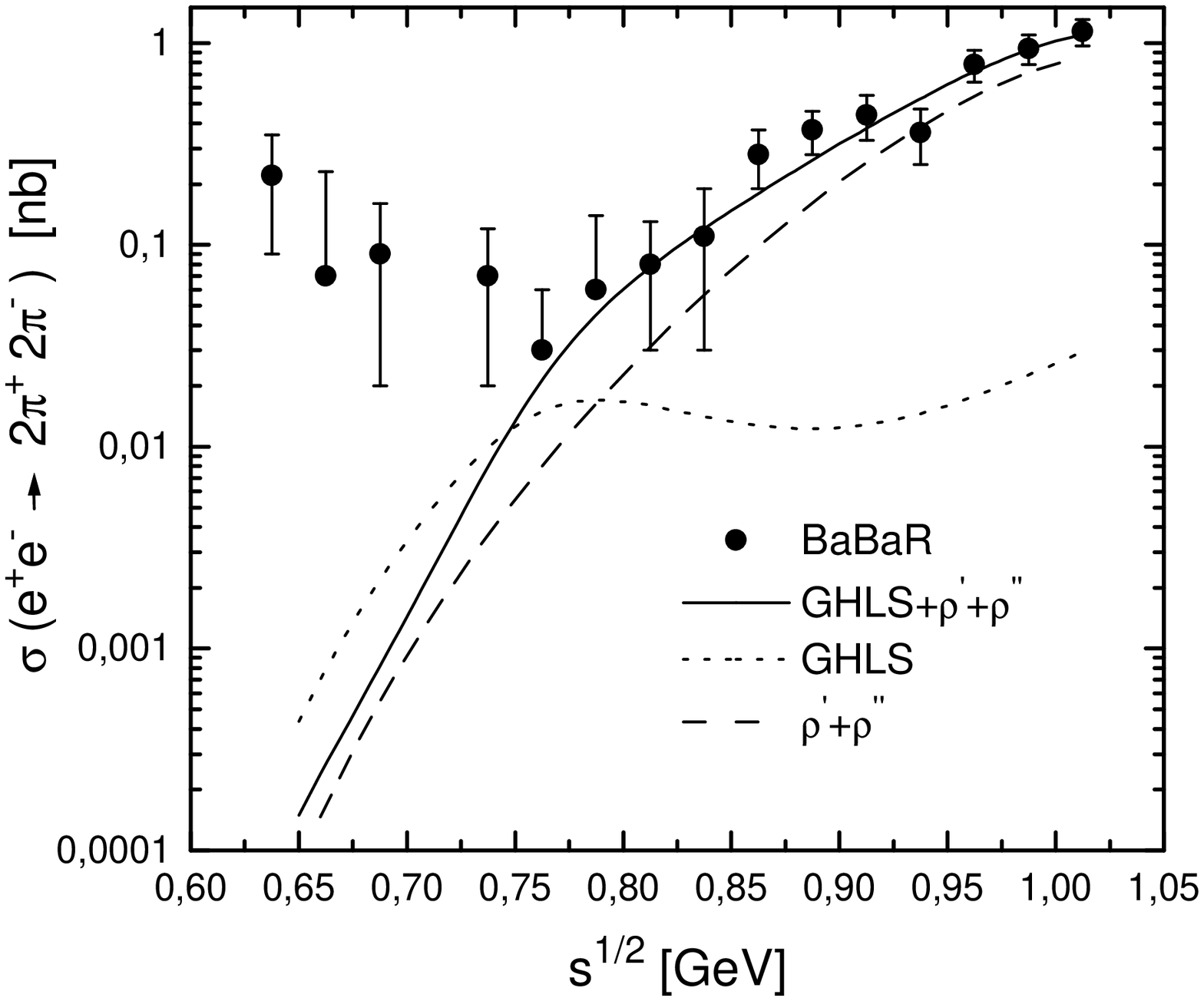}
  \caption{\label{fitbabar}Fitting BaBaR data \cite{babar}.}
\end{center}
  \end{figure}

The results of evaluation of the cross section of the reaction
$e^+e^-\to\pi^+\pi^-\pi^+\pi^-$ in the GHLS model with the
"canonical choice" (\ref{canon}) are shown in Fig.~\ref{fitcmd2}
and \ref{fitbabar} with the dotted line.
One can see that the simplest variant with the single $\rho(770)$
s-channel resonance cannot reproduce the data at
$\sqrt{s}\approx1$ GeV. Upon including the $\rho^\prime$,
$\rho^{\prime\prime}$ resonances with the couplings chosen by
analogy with the $\rho(770)$ ones one can improve the description
of the data \cite{cmd2,babar}. The results of fitting CMD-2 data
\cite{cmd2} and the BABAR ones \cite{babar} in the fixed width
approximation for the $\rho^\prime$, $\rho^{\prime\prime}$
resonances with the PDG masses and widths \cite{pdg} are shown in
Fig.~\ref{fitcmd2} and \ref{fitbabar}. It appears that at
$\sqrt{s}\approx1$ GeV the joint contribution of the
$\rho^\prime$, $\rho^{\prime\prime}$ resonances is by the factor
of thirty grater than the contribution of $\rho(770)$.

{\bf 6. Conclusion.} The heavier  axial vector resonances
$a^\prime_1$ and $a^{\prime\prime}_1$ contributions should be
added to the $a_1(1260)$ one in order to obtain the correct shape
of the spectrum in the decay $\tau\to3\pi\nu_\tau$. Similar
problem is found in the vector channel
$e^+e^-\to\pi^+\pi^-\pi^+\pi^-$. The contributions of heavier
resonances $\rho^\prime$ and $\rho^{\prime\prime}$ are required
for correct description of experimental data at energy
$\sqrt{s}\approx1$ GeV. However, contrary to the case of the
vector channel where additional contributions of $\rho^\prime$ and
$\rho^{\prime\prime}$ at the above energy exceed the $\rho(770)$
one, in the axial vector channel $\tau\to3\pi\nu_\tau$, each of
the additional contributions is smaller in magnitude  than the
contribution of pure GHLS with the single fitted $a_1$ resonance.
See Fig.~\ref{spec_const} and \ref{fixr_const}. But they
contribute almost coherently resulting in the acceptable shape of
the spectrum and the acceptable magnitude of the branching
fraction $B(\tau^-\to\pi^+\pi^-\pi^-\nu_\tau)\approx9\%$.


\end{document}